\documentclass[12pt,letterpaper]{article}
\usepackage{color}
\usepackage{hyperref}

\begin{document}

\hspace*{\fill} \href{https://creativecommons.org/licenses/by-nc-sa/4.0/}{\bf CC BY-NC-SA 4.0} 

\vspace*{1cm}

\begin{center}

{\huge \bf  How to make your research group \\[0.2cm] more inclusive for autistic trainees\footnote{Throughout this guide I use identity-first language (``autistic person'') because it is strongly preferred by a large majority of autistic adults (see, e.g., A. Taboas et al., Autism 27(2), 565 (2023), \href{https://doi.org/10.1177/13623613221130845}{DOI: 10.1177/13623613221130845}).  When speaking to an autistic person, ask them what their own personal terminology preference is, and use it.}\footnote{A free printable poster is available as an ancillary file at \href{https://arXiv.org/abs/2410.17929}{https://arXiv.org/abs/2410.17929}.}
}

\end{center}

\vspace*{2cm}

\noindent
As a university professor, one of my most important responsibilities is mentoring the junior members of my research group and creating an inclusive environment in which they can thrive.  Since my autism diagnosis two years ago, colleagues have asked me how they can make their research groups more welcoming to autistic trainees.  This short guide, based on conversations with autistic students and academics, intense reflection on my own lived experience, and a deep dive into the literature, provides five concrete steps toward this goal. \\

\hspace*{\fill} {\it Prof.\ Heather E.\ Logan, Physics Department, Carleton University, October 2024}

\vspace*{\fill}

\noindent
{\it I thank my friends and colleagues who have contributed their perspectives and feedback during the development of this guide, including Luke Beardon, Adrian Chan, Sarah Finn, Lorna Jarrett, Christine A.~Jenkins, Ira Kraemer, Tess LeBlanc, Cathy Malcolm Edwards, Keivan Stassun, Daniel Stolarski, Rowan Thomson, and Brandon G.\ Villalta Lopez. This guide was written on the traditional and unceded territory of the Algonquin Anishinabeg people.}

\vspace*{1cm}

\noindent

\pagebreak

\noindent
{\Large \bf 1. Dispel your misconceptions.}
\vspace{1cm}

Comprising an estimated 3--4\% of the population [1,2], autistic people are increasingly recognized as an essential facet of natural human neurodiversity [3].  Autism becomes a disability when the social and physical environment is set up solely to accommodate the needs and preferences of the non-autistic majority.  Despite pathologizing narratives [4,5], caricatured portrayals in popular media, and pervasive unconscious bias [6], most autistic people can and do thrive in a full range of careers---including academia [7]---when our needs are met [8]. \\

Autism is genetic~[9] and is lifelong.  Women are no less likely to be autistic than men are [1], though persistent gender bias in referrals results in more than 75\% of autistic girls remaining undiagnosed by age 18 [1]; more than 80\% of all autistic adults over age 30 also remain undiagnosed [2].  Those who do know themselves to be autistic may be reluctant to disclose due to fear of stigma or of being disbelieved [10].  Establishing an autistic-friendly research environment is thus important whether or not anyone in your group has disclosed their autism to you.  This begins with recognizing that each person's needs are unique, and cannot be guessed based on your prior knowledge or experiences.  Proactively building flexibility into your workplace is a key principle of universal design and benefits multiple groups.  \\[1cm]

\hrule 
\hspace*{0.1cm}
{\small 

[1] R. McCrossin, ``Finding the true number of females with autistic spectrum disorder by estimating the biases in initial recognition and clinical diagnosis,'' Children 9(2), 272 (2022). \href{https://doi.org/10.3390/children9020272}{DOI: 10.3390/} \href{https://doi.org/10.3390/children9020272}{children9020272}

[2] E. O'Nions et al., ``Autism in England: assessing underdiagnosis in a population-based cohort study of prospectively collected primary care data," The Lancet Regional Health Europe 29, 100626 (2023). \href{https://doi.org/10.1016/j.lanepe.2023.100626}{DOI: 10.1016/j.lanepe.2023.100626}

[3] See, e.g., E. Pellicano et al., ``Annual Research Review: Shifting from `normal science' to neurodiversity in autism science," Journal of Child Psychology and Psychiatry 63(4), 381 (2021). \href{https://doi.org/10.1111/jcpp.13534}{DOI: 10.1111/jcpp.13534}

[4] E. Loughran, ``The rampant dehumanization of autistic people,'' NeuroClastic (online), 6 Jan 2020. \href{https://neuroclastic.com/the-rampant-dehumanization-of-autistic-people/}{https://neuroclastic.com/the-rampant-dehumanization-of-autistic-people/}

[5] M. Botha et al., `` `Autism research in crisis': a mixed method study of researchers' constructions of autistic people and autism research," Frontiers in Psychology 13, 1050897 (2022). \href{https://doi.org/10.3389/fpsyg.2022.1050897}{DOI: 10.3389/} \href{https://doi.org/10.3389/fpsyg.2022.1050897}{fpsyg.2022.1050897} 

[6] L. N. Praslova, ``Autism doesn't hold people back at work. Discrimination does.,'' Harvard Business Review (online), 13 Dec 2021. \href{https://hbr.org/2021/12/autism-doesnt-hold-people-back-at-work-discrimination-does}{https://hbr.org/2021/12/autism-doesnt-hold-people-back-at-} \href{https://hbr.org/2021/12/autism-doesnt-hold-people-back-at-work-discrimination-does}{work-discrimination-does} 

[7] S. C. Jones, ``Autistics working in academia: what are the barriers and facilitators?," Autism 27(3), 822 (2022) \href{https://doi.org/10.1177/13623613221118158}{DOI: 10.1177/13623613221118158}; 
``Advice for autistic people considering a career in academia," Autism 27(7), 2187 (2023). \href{https://doi.org/10.1177/13623613231161882}{DOI: 10.1177/13623613231161882} 

[8] C. Morgan et al., ``Enabling NeuroDiverse Inclusive Scientific Careers," Centre for Employment, Work and the Professions, Institute of Life and Earth Sciences, Heriot-Watt University (2022). \href{https://disc.hw.ac.uk/en-disc-enabling-neurodiverse-inclusive-science-careers/}{https://disc.hw.ac.uk/en-disc-enabling-neurodiverse-inclusive-science-careers/}

[9] S. Sandin et al., ``The heritability of autism spectrum disorder,'' JAMA 318(12), 1182 (2017). \href{https://doi.org/10.1001/jama.2017.12141}{DOI: 10.1001/jama.2017.12141}

[10] A. Farsinejad et al., ``Autism disclosure - the decisions autistic adults make," Research in Autism Spectrum Disorders 93, 101936 (2022). \href{https://doi.org/10.1016/j.rasd.2022.101936}{DOI: 10.1016/j.rasd.2022.101936}
}

\pagebreak

\noindent
{\Large \bf 2. Communicate clearly.}
\vspace{1cm}

While autism is often framed as a communication disorder, it is more accurately described as a mismatch of communication styles~[1].  You can bridge the communication divide by applying cross-cultural communication skills even when you think your trainees share your own culture, and by adapting your communication style to match that of your trainees as needed.  This benefits not only autistic and other neurodivergent trainees, but also international, first-in-family, and other equity-seeking students who may be unaware of some aspects of academic culture~[2]. In particular, if you think a trainee is doing good work and should apply for a scholarship, job, graduate program, or conference, tell them! They may undervalue their accomplishments without direct feedback~[3]. \\

Be explicit and straightforward in your communications~[4] and avoid reliance on metaphors, colloquialisms, or unspoken messages.  Provide critical information in writing so that trainees can refer to it as needed---for example, shared note-taking documents can boost group productivity while helping you keep track of trainee progress.  Provide direct instruction on research practices like record-keeping, but allow trainees to adapt their systems to fit their needs.  Provide timelines and context: trainees are less likely to misinterpret instructions when they understand how each step fits into the bigger picture, leading to enhanced ownership of their contributions and fewer costly mistakes.  Meet regularly with trainees and be clear about how and when they can contact you.  Listen carefully and take all questions seriously, especially when the answer seems obvious to you.  Autistic trainees may be afraid to ask for clarification more than once due to prior negative experiences~[2]; make a habit of checking for understanding while recognizing that autistic people may need more time to process information.  Clear communication benefits all researchers!
\\[1cm]

\hrule
\hspace*{0.1cm}
{\small

[1] D. E. M. Milton, ``On the ontological status of autism: the `double empathy problem'," Disability \& Society 27(6), 883 (2012). \href{https://doi.org/10.1080/09687599.2012.710008}{DOI: 10.1080/09687599.2012.710008}

[2] ``Inclusivity and universal design," Employment Autism (online; accessed 26 Jul 2024). \\ \href{https://employmentautism.org.uk/tools-and-resources/inclusivity-and-universal-design/}{https://employmentautism.org.uk/tools-and-resources/inclusivity-and-universal-design/}

[3] C. M. Syharat et al., ``Experiences of neurodivergent students in graduate STEM programs,'' Front. Psychol. 14, 1149068 (2023). \href{https://doi.org/10.3389/fpsyg.2023.1149068}{DOI: 10.3389/fpsyg.2023.1149068}

[4] Work Inclusion Project, Warwick University Social Inclusion Group, ``Autism and communication guidance" (2022). \href{https://warwick.ac.uk/services/socialinclusion/projects/letstalkaboutdisability/autism/2022_twip_autism_and_communication.pdf}{https://warwick.ac.uk/services/socialinclusion/projects/letstalkaboutdisability/} \href{https://warwick.ac.uk/services/socialinclusion/projects/letstalkaboutdisability/autism/2022_twip_autism_and_communication.pdf}{autism/2022\_twip\_autism\_and\_communication.pdf}
}

\pagebreak

\noindent
{\Large \bf 3. Check the sensory environment.}
\vspace{1cm}

Differences in sensory processing are common in the autistic population~[1].  Background noise, lighting, temperature, smells, textures, and visual distractions can lead to sensory overwhelm and prevent an autistic trainee from doing their best work~[2].  Frequent interruptions, even small ones, can disrupt autistic thought processes, sapping energy and reducing focus~[3].  Unstructured networking events and crowded poster sessions can be overwhelming to some autistic people~[4]; instead of relying solely on them for recruitment, promote alternative methods such as centralized departmental publicity for research group openings with clear instructions on how to apply.  \\

Each person's sensory profile is unique, and autistic people may be unaware that their sensory experiences differ from others'.  Encourage trainees to experiment with different work environments and identify the conditions under which they function best.  Many sensory issues are easy to accommodate by allowing use of earplugs or adjustable task lighting, adjusting seating arrangements and social expectations, scheduling quiet breaks between high-intensity activities such as meetings or social events, establishing explicit turn-taking or alternative contribution modes (e.g., written) in group meetings, and permitting trainees to work from home or use videoconferencing when their physical presence on site is not required~[4].   Be alert to environments or situations in which a trainee disengages or becomes distressed, but be aware that the toll of sensory overwhelm may appear only hours after the offending exposure; a trainee missing workdays may be a sign of problems with the sensory or social environment.  See~[2] for specific examples of sensory sensitivities and accommodations.  \\[1cm]

\hrule
\hspace*{0.1cm}
{\small

[1] See, e.g., L. Crane et al., ``Sensory processing in adults with autism spectrum disorders," Autism 13 (3), 215 (2009). \href{https://doi.org/10.1177/1362361309103794}{DOI: 10.1177/1362361309103794}

[2] L. Praslova, ``Sensory safety: a must of neurodiversity inclusion in the workplace," Specialisterne (online), 18 Jan 2023. \href{https://ca.specialisterne.com/sensory-safety-a-must-of-neurodiversity-inclusion-in-the-workplace/}{https://ca.specialisterne.com/sensory-safety-a-must-of-neurodiversity-inclusion-} \href{https://ca.specialisterne.com/sensory-safety-a-must-of-neurodiversity-inclusion-in-the-workplace/}{in-the-workplace/}

[3] F. Murray, ``Me and monotropism: a unified theory of autism," The Psychologist (online), 30 Nov 2018. \href{https://www.bps.org.uk/psychologist/me-and-monotropism-unified-theory-autism}{https://www.bps.org.uk/psychologist/me-and-monotropism-unified-theory-autism}

[4] See, e.g., T. Feder, ``Community heightens attention to accessibility for physicists with disabilities," Physics Today 76(3), 22 (2023). \href{https://doi.org/10.1063/PT.3.5194}{DOI: 10.1063/PT.3.5194}
}

\pagebreak 

\noindent
{\Large \bf 4. Be aware of different cognitive profiles.}
\vspace{1cm}

Autism is fundamentally a different way of thinking~[1].  Autistic people often take wildly different approaches to problems, and notice different details and patterns, than non-autistic people: embracing this diversity of thought within your research group promotes better decision-making and more creative approaches to problem-solving.  Autistic thought processes tend to be ``bottom-up," collecting details first and then combining them to build a coherent bigger picture~[2].  Facilitate this by providing concrete examples to illustrate abstract concepts and give additional detail when asked so that trainees can incorporate novel information into their mental schema before moving on~[3].  Autistic people can be ``deep divers'' when researching a topic: capitalize on this when appropriate, but provide clear guidance on how much depth and scope is required and how trainees should prioritize their time expenditures.  \\

Bottom-up thinking requires time and effort: allow sufficient processing time, especially when trainees are expected to make decisions, and avoid requiring multitasking where possible~[1].   Autistic people tend to do their best work in familiar and predictable environments: unavoidable changes to schedules, processes, or physical settings should be communicated at least a day in advance, with an explanation of their necessity, to allow for pre-processing.  Provide group meeting agendas in advance so that trainees can come prepared.  Allow trainees to familiarize themselves in advance with the location and equipment for high-stakes activities such as presentations or thesis defences, and encourage them to keep extraneous distractions (such as unfamiliar clothing) to a minimum during these events.  Consider developing a ``buddy system'' by which group members can help each other navigate unfamiliar environments during conference travel or fieldwork.  Be aware that autistic people typically have a much more uneven skill-set than the majority: a single person can be a top performer in some areas while struggling in others~[4].  \\[1cm]

\hrule
\hspace*{0.1cm}
{\small

[1] F. Murray, ``Me and monotropism: a unified theory of autism," The Psychologist (online), 30 Nov 2018. \href{https://www.bps.org.uk/psychologist/me-and-monotropism-unified-theory-autism}{https://www.bps.org.uk/psychologist/me-and-monotropism-unified-theory-autism}

[2] E. Sanborne, ``What is bottom-up thinking in autism?," Autistic PhD blog (online), 7 Aug 2023. \href{https://autisticphd.com/theblog/what-is-bottom-up-thinking-in-autism/}{https://autisticphd.com/theblog/what-is-bottom-up-thinking-in-autism/}

[3] See, e.g., Emma, ``How I learn\ldots \ and the difficulties this causes," Undercover Autism blog (online), 13 Aug 2020. \href{https://undercoverautism.org/2020/08/13/how-i-learn-and-the-difficulties-this-causes/}{https://undercoverautism.org/2020/08/13/how-i-learn-and-the-difficulties-this-causes/}

[4] See, e.g., Oolong (F. Murray), ``Autistic skill sets: a spiky profile of peaks and troughs," NeuroClastic (online), 5 Jul 2019. \href{https://neuroclastic.com/autistic-skill-sets/}{https://neuroclastic.com/autistic-skill-sets/}
}

\pagebreak

\noindent
{\Large \bf 5. Model inclusivity to your group.}
\vspace{1cm}

Autistic people naturally display different body language, facial expressions, and vocal rhythm and tone than the majority of the population~[1].  This leads most non-autistic people to form negative first impressions and be less likely to pursue interactions with autistic people than with members of their own neurotype~[2].  It also leads to frequent misinterpretations of autistic people's intent and emotional state, with potentially traumatizing consequences~[3].  Most autistic people attempt to mitigate this by effortfully ``masking" or camouflaging their autistic traits.  Masking consumes a tremendous amount of cognitive energy and contributes to social isolation, burnout, and mental health problems~[4,5]. \\

Your trainees will do their best work when they feel safe to be themselves.  As leader, you set the tone for your group: model acceptance of physical and vocal autistic traits as well as all other forms of diversity, and do not tolerate bullying or mockery.  Recognize that you hold immense power over your trainees' futures~[4] and do everything you can to identify sources of trainee anxiety and to reduce or eliminate them~[6].  Encourage group social relations but avoid imposing forced socialization; autistic sociality can look very different from non-autistic norms.  Model direct, straightforward, non-judgemental communication: state your feelings, and ask for clarification rather than making assumptions about others' emotional states~[3,7].  
Actively structure your training environment to leverage each trainee's strengths while accommodating their weaknesses, and establish a clear and easily-navigated process for implementing accommodations for co-occurring disabilities~[8]. Critically examine your ``best practices'' to ensure that they really are ``best'' for everyone.  \\[1cm]

\hrule
\hspace*{0.1cm}
{\small

[1] D. Bercovici, ``Autistic speech \& nonverbal communication differences," Embrace Autism (online), 16 Jun 2023. \href{https://embrace-autism.com/autistic-verbal-and-nonverbal-communication-differences/}{https://embrace-autism.com/autistic-verbal-and-nonverbal-communication-differences/}

[2] N. J. Sasson et al., ``Neurotypical peers are less willing to interact with those with autism based on thin slice judgments," Scientific Reports 7, 40700 (2017).  \href{https://doi.org/10.1038/srep40700}{DOI: 10.1038/srep40700}

[3] See, e.g., Work Inclusion Project, Warwick University Social Inclusion Group, ``Autism and communication guidance" (2022). \href{https://warwick.ac.uk/services/socialinclusion/projects/letstalkaboutdisability/autism/2022_twip_autism_and_communication.pdf}{https://warwick.ac.uk/services/socialinclusion/projects/letstalkabout} \\ \href{https://warwick.ac.uk/services/socialinclusion/projects/letstalkaboutdisability/autism/2022_twip_autism_and_communication.pdf}{disability/autism/2022\_twip\_autism\_and\_communication.pdf}

[4] C. M. Syharat et al., ``Experiences of neurodivergent students in graduate STEM programs,'' Front. Psychol. 14, 1149068 (2023). \href{https://doi.org/10.3389/fpsyg.2023.1149068}{DOI: 10.3389/fpsyg.2023.1149068}

[5] S. L. Field et al., ``A meta-ethnography of autistic people's experiences of social camouflaging and its relationship with mental health," Autism 28(6), 1328 (2024). \href{https://doi.org/10.1177/13623613231223036}{DOI: 10.1177/13623613231223036}

[6] L. Beardon, {\it Avoiding Anxiety in Autistic Adults: A Guide for Autistic Wellbeing}. Sheldon Press, 2021. \href{https://www.sheldonpress.co.uk/titles/luke-beardon/avoiding-anxiety-in-autistic-adults/9781529394740/}{ISBN-13: 9781529394740}.

[7] I. Kraemer, ``Neurotypicals: Listen to Our Words, Not Our Tone,'' Autistic Science Person blog (online), 9 Jan 2021. \href{https://autisticscienceperson.com/2021/01/09/neurotypicals-listen-to-our-words-not-our-tone/}{https://autisticscienceperson.com/2021/01/09/neurotypicals-listen-to-our-} \href{https://autisticscienceperson.com/2021/01/09/neurotypicals-listen-to-our-words-not-our-tone/}{words-not-our-tone/}

[8] J. Rudd, ``What universities can do to support their autistic employees,'' Times Higher Education (online), 27 Oct 2022. \href{https://www.timeshighereducation.com/campus/what-universities-can-do-support-their-autistic-employees}{https://www.timeshighereducation.com/campus/what-universities-can-do-} \href{https://www.timeshighereducation.com/campus/what-universities-can-do-support-their-autistic-employees}{support-their-autistic-employees}
}

\end{document}